\documentclass[11pt]{article}
\usepackage{graphicx}
\usepackage{amsmath}
\usepackage{amssymb}
\usepackage{amsxtra}

\title{Gravitational waves in the modified gravity}
\author{Sourav Roy Chowdhury\\Research Institute of Physics, Southern Federal University,\\ 344090 Rostov on Don, Russia.,\\ Email : roic@sfedu.ru\\
Maxim Khlopov\\Research Institute of Physics, Southern Federal University,\\ 344090 Rostov on Don, Russia.\\Université de Paris,
CNRS, Astroparticule et Cosmologie,\\ F-75013 Paris, France.\\ Center for Cosmoaprticle Physics Cosmion,\\ National ResearchNuclear University “MEPHI”,\\ 31 Kashirskoe Chaussee, 115409 Moscow, Russia\\Email : khlopov@apc.in2p3.fr\\}

\begin{document}
\maketitle

\begin{abstract}
We have taken a modified version of the Einstein Hilbert action, $ f(R, T^\phi) $ gravity under consideration, where $T^\phi$ is the energy-momentum tensor trace for the scalar field under consideration. The structural behaviour of the scalar field considered varies with the form of the potential. The number of polarization modes of gravitational waves in modified theories has been studied extensively for the corresponding fields. There are two additional scalar modes, in addition to the usual two transverse-traceless tensor modes found in general relativity: a massive longitudinal mode and a massless transverse mode (the breathing mode).
\end{abstract}


% optionally
%\noindent PACS: ... list of PACS codes

\section{Introduction}\label{s:intro}

The FLRW metric is an exact solution to Einstein's equations, achieved under the implication of space homogeneity and isotropy. It has been well recognized for satisfactorily explaining several other observational evidence about our Universe, including the distribution of large-scale galaxies and the near-uniformity of the CMB temperature \cite{Planck}. The FLRW metric \cite{cao} underpins the existing accepted cosmological model, which is quite good at likely fitting continued application data sets and trying to explain measured cosmic acceleration. The fact that the cosmological space-time metric differs from the FLRW metric would have massive consequences for inflation theory as well as fundamental physics.

Alternative explanations of gravity have long been considered to prevent a few of the contradictions in conventional cosmology \cite{Clifton,paddy}. A potential substitute is the $f(R, T)$ gravity, created recently by Harko et al. \cite{harko}. The latest identification of gravitational waves (GWs) by the Advanced LIGO group has opened up a massive door to analyse the Universe. \cite{abb1,abb2,abb3}. Apart from directly detecting GWs with LIGO/VIRGO interferometers, one could use the informal identification of GWs by assessing the substantial reduction of the orbital period of stellar binary configuration. Detecting nano-Hertz GWs with a pulsar timing array includes timing various millisecond pulsars, which seem to be extremely stable celestial clocks, according to Jenet \cite{jenet}. This connection is effected by the angular distance ($\theta$) between both the two pulsars, as well as the polarization of GW and graviton mass, according to $C(\theta)$ \cite{lee}.

The range of the GW, including its polarization modes, is based on the theories. In the radiative domain, the polarization and dispersion of GWs in vacuum are two critical features of GWs that distinguish between the authenticity of gravity theories.GWs can also have up to six conceivable polarization states in substitute metric theories, four more than GR permits. 

Hou et al. \cite{hou} carried out a detailed analysis of the polarization mode for the Horndeski theory. Using GWs polarization, Alves et al. \cite{alves1} investigated the f(R) framework. In $ f (R) $ gravity metric methodology, the model, including other $ f (R) $ theoretical models, confirms the effectiveness of scalar degrees of freedom. There is a scalar mode of polarization of GWs exists in theory. This polarization mode appears in two different states: a massive longitudinal mode and a transverse massless breathing mode with non-vanishing trace \cite{gogoi}.  Capozziello and Laurentis 
\cite{Capozziello} find the palatini formalism, conformal transformations and find the new polarization states for gravitational radiation for the higher order of extended gravity $ ( f(R) = R + \alpha R^2) $ Later on, Alves et al. \cite{Alves2} studied for $ f (R, T) $ and $ f (R, T^{\phi}) $ theoretical models, .

In this article, we studied the polarization modes based on the potential, which is a function of the scalar field under the framework of modified gravity $ f (R, T^{\phi}) $ for the vacuum system. In Sec. \ref{s1} we developed the basic formalism of the modified gravity. The scalar field structure and equation of motion is developed in Sec. \ref{s2}. Polarization modes using Newman-Penrose (NP) formalism is analyzed Sec. \ref{s4}. And in Sec. \ref{s5} we conclude the results.

\section{Basic formalism of the modified gravity} \label{s1}

In the context of modified gravity \cite{harko}, for the vacuumed system, the total action including the scalar field can be introduced in the following manner,

\begin{equation}
S = \int d^4 x \sqrt{-g} \big[f(R,T^{\phi})+
\mathcal{L(\phi, \partial_{\mu} \phi)} \big],\label{3}
\end{equation}

where $ R $ stands for the Ricci scalar, while $ T^{\phi} $ is the trace of the scalar field's energy-momentum tensor.

The field's action, with g as the metric's determinant and signature (-, +, +, +).
We use geometric units with the formula $G = c = 1$.

Following that, we considered $\mathcal{L(\phi, \partial_{\mu} \phi)} = \mathcal{L}_{\phi}$. Here $ \mathcal{L}_{\phi} $ is the standard Lagrangian density for a real scalar field ($ \phi $), as follow  \cite{moraes},  

\begin{equation}
\mathcal{L}_\phi = \frac{1}{2} \nabla_\alpha \phi \nabla^\alpha \phi -V(\phi). \label{4}
\end{equation}

A self-interacting potential is represented by $V(\phi)$.
In this theory, matter fields have a relatively limited coupling to gravity and no coupling to the scalar field.

The stress-energy tensor can define as 

\begin{equation}
T^\phi_{\mu \nu} = -\frac{2}{\sqrt{-g}} \frac{\delta (\sqrt{-g} \mathcal{L} )}{\delta g^{\mu \nu}}.
\end{equation}

 We assumed that the Lagrangian density $ L $ is free of its derivatives and is only conditional on the metric tensor modules  $g^{\mu \nu}$. 

Therefore, the energy-momentum tensor of the scalar field is 

\begin{equation}
T^\phi_{\mu \nu} = \frac{1}{2}g_{\mu \nu} \nabla_\alpha \phi \nabla^\alpha \phi -g_{\mu \nu} V(\phi)-\nabla_\mu \phi \nabla_\nu \phi, \label{5}
\end{equation}

and the corresponding trace is given by

\begin{eqnarray}
T^\phi=\nabla_\alpha \phi \nabla^\alpha \phi - 4 V(\phi). \label{6}
\end{eqnarray}

The generalized form of the Einstein field equation in vacuum in the involvement of scalar field is obtained by varying the gravitational field's action S concerning the metric tensor components, $g_{\mu \nu}$, and then on integration as follow,

\begin{equation}
f_R R_{\mu \nu}-\frac{f}{2}g_{\mu \nu}= \frac{1}{2}T_{\mu \nu}^\phi +f_{T} T_{\mu \nu}^\phi -f_{T} g_{\mu \nu}\mathcal{L}_\phi\label{9}
\end{equation}

Here, $f_R = f_R(R,T^\phi)$ and $f_T = f_T(R,T^\phi)$ denotes $\partial f(R,T^\phi)/ \partial R$ and $\partial f(R,T^\phi)/ \partial T^\phi$, respectively.

We assume that the modified gravity function $f(R,T^{\phi})$ is given by $f(R,T^{\phi})= R + \beta T^{\phi}$, $\beta$ is an arbitrary constant. The field equation immediately takes the following form,
\begin{equation}
G_{\mu\nu} = \frac{1}{2}[T_{\mu\nu}^{\phi}+g_{\mu\nu}\beta T^{\phi}-2\beta \nabla_{\mu}\phi\nabla_{\nu}\phi].\label{10}
\end{equation}

\section{Scalar Field}\label{s2}

On contraction and simplification, the Eq. (\ref{9}) the Ricci scalar of can be obtained as follows,
\begin{eqnarray}
R = -\frac{1}{2 }[ 4 \beta T^{\phi} +T^\phi -2\beta\nabla_\mu \phi \nabla^\mu \phi]\label{11}
\end{eqnarray}   

The equation of motion for the scalar field can be found from the covariant divergence of the field Eq. (\ref{10})  as follows,
\begin{eqnarray}
(1+2 \beta)\Box \phi + (1+4\beta)\bigg(\frac{\partial V}{\partial \phi}\bigg) = 0.\label{12}
\end{eqnarray}

Since we are considering the vacuum system, we consider the potential in the following form,
\begin{equation}
V(\phi) = \frac{1}{2} \mu^2 \phi^2 + \frac{1}{4} \lambda \phi^4, \label{18}
\end{equation}
where, $\mu \textrm{~and~} \lambda$ are real constants.

\begin{figure}[h!]\centering
	\includegraphics[width=8cm]{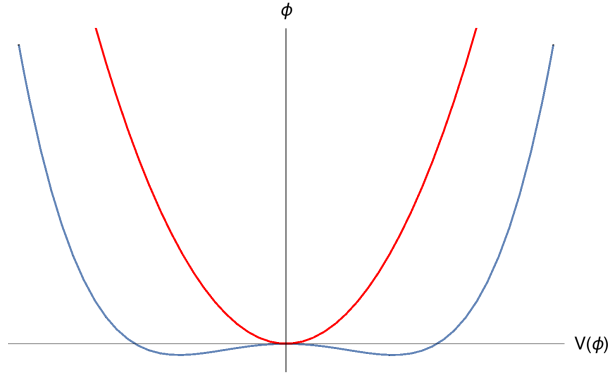}
	\caption{ variation of the potential $V({\phi})$ with scalar variable $ \phi $. Red curve shows the variation for $ \mu^2 >0 $, and Blue curve shows the variation for $ \mu^2 < 0 $. } \label{f2}
\end{figure}

We limited ourselves to first-order terms in $\phi$.
The third term of Eq. (\ref{12}) disappear as a result of this estimation. $ V $ is being expanded around the non-null minimum value $ V_0 $. $\eta = \phi-\phi_0$ can be used to expand the field.
Following identical approach as before, we encounter it with such assumptions and first-order restrictions, 

\begin{equation}
\Box \phi + \Bigg(\frac{1+4 \beta }{1+2  \beta}\Bigg)\bigg(\frac{\partial V}{\partial \phi}\bigg) = 0. \label{19}
\end{equation}  

The field equations in the linear region has been investigated and leads to the  solution in the following form,
\begin{equation}
\phi(x) = \phi' + \phi_1 \exp{(iq_\rho x^\rho)},\label{15}
\end{equation}

Solution of the scalar field corresponds to the above equation can be written as in Eq. (\ref{15}) with,
\begin{eqnarray}
\phi' = \phi_0 - \Bigg(\frac{\mu^2 + \lambda \phi_0^2}{\mu^2 + 3 \lambda \phi_0^2 }\Bigg)\phi_0,
\end{eqnarray}  

and
\begin{equation}
q_\mu q^\mu = (\mu^2+3\lambda \phi_0^2) \Bigg(\frac{1+4 \beta }{1+2 \beta}\Bigg). \label{14}
\end{equation}

\begin{figure}[h!]\centering
	\includegraphics[width=10cm]{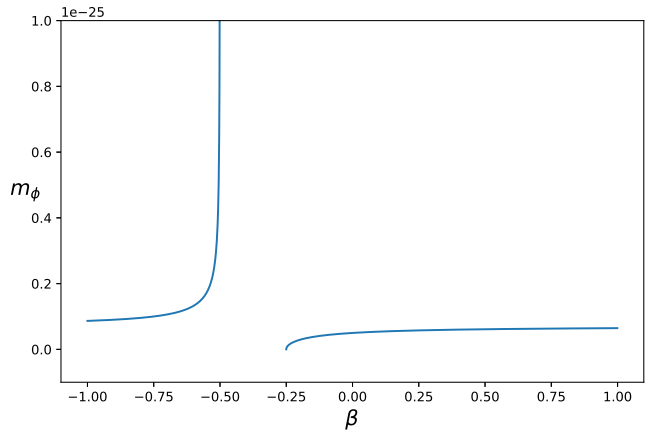}
	\caption{ variation of the mass function $m_{\phi}$ with coupling constant $ \beta $.} \label{f1}
\end{figure}

The variation of the effective mass ($ m_{\phi} $) with the coupling constant is shown in the Fig. \ref{f1}. The restricted range is for $ -0.50 \leq \beta \leq -0.25 $, from Eq. (\ref{14}).

Corresponding energy of the system can be written as
\begin{equation}
E = \pm \Bigg[ q^2 + (\mu^2+3\lambda \phi_0^2) \Bigg(\frac{1+4 \beta }{1+2 \beta }\Bigg) \Bigg]^{1/2}
\end{equation}

\begin{figure}[h!]\centering
	\includegraphics[width=9cm]{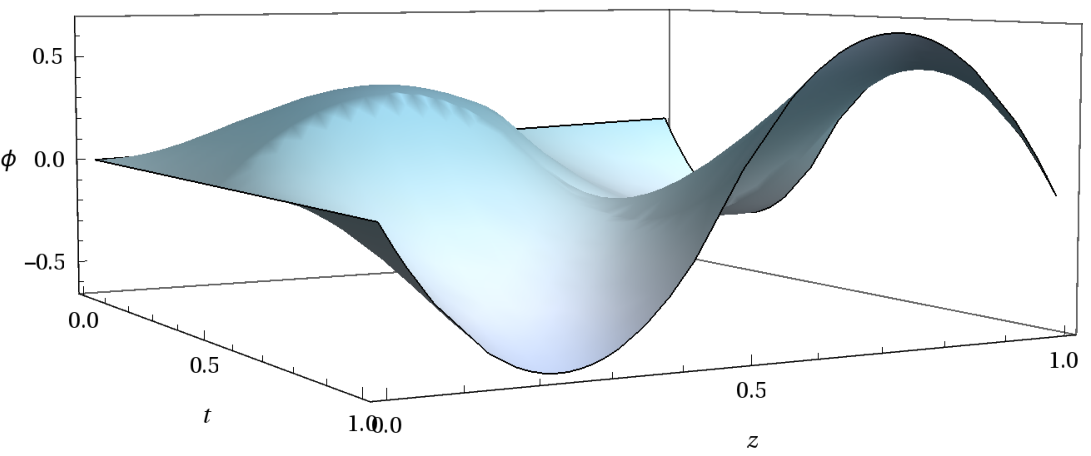}\\
	\includegraphics[width=9cm]{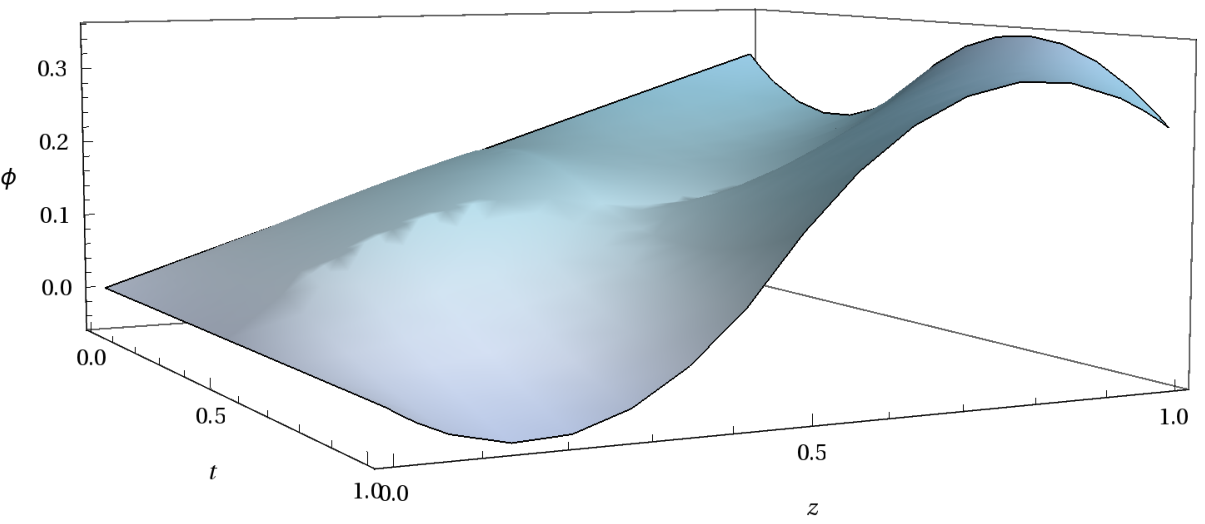}
	\caption{Propagation for the perturbation of vacuum scalar field. Upper panel shows the variation for $ \mu^2<0 $, and lower panel shows the variation for $ \mu^2>0 $. Considered $ \beta  = 0.5. $}
\end{figure}

The first-order minimally coupled scalar field exposes an effective cosmological constant, as follows:

\begin{equation}
\Lambda =  \frac{V_0}{2} \bigg(4 \beta + 1 \bigg).
\end{equation}

With $\lambda$ being a positive constant, the potential in Eq. (\ref{18}) could be categorized into two situations: (i) $\mu^2 > 0$, and (ii) $\mu^2<0$.
This is what the universe needs to be stable.
While the minimum scalar field for $\mu^2 < 0$ is non-zero, the effective cosmological constant is non-zero.
The cosmological constant that is effective is 

\begin{equation}
\Lambda =  -\frac{1}{2} \bigg[ \beta \bigg(\frac{\mu^4}{ \lambda}\bigg) + \frac{\mu^4}{4 \lambda} \bigg].
\end{equation}

The steady minimum of the scalar field is zero for $\mu^2 > 0$, which causes the effective cosmological constant ($ \Lambda $) to be zero.

\section{Polarization modes of the modified gravity}\label{s4}

\subsection*{Newman-Penrose formalism} 

The Newman-Penrose (NP) \cite{New1,New2} method is used to find additional polarization modes; further information is available in the references \cite{eardley1,eardley2}. Tetrads are a combination of standardized linearly independent vectors $(e_t, e_x, e_y, e_z)$ that could be used to describe the NP quantities that correlate to all of the six polarization modes of GWs at any spatial position. The NP tetrads $k,~l,~m~,\bar{m}$. can be used to recognize these vectors. The actual null vectors are as follows:

\begin{eqnarray}
k = \frac{1}{\sqrt{2}}(e_t+e_z),~~
l = \frac{1}{\sqrt{2}}(e_t-e_z),
\end{eqnarray}

And the other two complex null vector are,  
\begin{eqnarray}
& & m = \frac{1}{\sqrt{2}}(e_x+ie_y),~~
\bar{m}= \frac{1}{\sqrt{2}}(e_x-ie_y).\\
& & -k.l=m. \bar{m}=1, ~~E_a = (k, l, m, \bar{m}).\nonumber
\end{eqnarray}
While all other dot product vanishes.

In the NP notation, the indefinable components of the Riemann tensor $R_{\lambda \mu \kappa \nu}$ are defined by ten components of the Wely tensor ($\Psi$'s), nine components of the traceless Ricci tensor ($\Phi$'s), and a curvature scalar ($\Lambda$). They are reduced to six by some symmetrical and differential properties: $\Psi_2, \Psi_3,\Psi_4 \textrm{~and~} \Phi_{22}$ are real and  $ \Psi_3\textrm{~and~}\Psi_4 $ are complex. These NP variables are associated with the following components of the Riemann tensor in the null tetrad basis: 

\begin{eqnarray}
& &\Psi_2 = -\frac{1}{6} R_{lklk} \sim  \textrm{longitudinal scalar mode,}\nonumber\\
& &\Psi_3 = -\frac{1}{2} R_{lkl \bar{m}} \sim  \textrm{vector-x \& vector-y modes,}\nonumber\\
& &\Psi_4 = - R_{l\bar{m}l\bar{m}} \sim  \textrm{+,} \times \textrm{tensorial mode,}\nonumber\\
& &\Phi_{22} = - R_{lml\bar{m}} \sim  \textrm{breathing scalar mode.} \label{e30}
\end{eqnarray} 

The additional nonzero NP variables are $\Phi_{11} = 3 \Psi_2 / 2$, $\Phi_{12}=\Phi_{21}= \Psi_3$ and $\Lambda = \Psi_2/2$, respectively. All of them can be defined base on the variables in Eq. (\ref{e30}).

The group E(2), the group of the Lorentz group for massless particles, can be used to classify these four NP variables $\Psi_2, \Psi_3,\Psi_4, \textrm{~and~}  \Phi_{22}$  based on their transformation properties. Only $ \Psi_2 $ is invariant, and the amplitudes of the four NP variables are not observer-independent, according to these transformations. The absence (zero amplitude) of some of the four NP variables, on the other hand, is not dependent on the observer. 

The following relations for the Ricci tensor and the Ricci  scalar hold:
\begin{eqnarray}
& & R_{lklk}= R_{lk},\nonumber \\
& & R_{lklm}= R_{lm},\nonumber\\
& & R_{lkl \bar{m}} = R_{l \bar{m}},\nonumber\\
& & R_{l \bar{m} l \bar{m}}= \frac{1}{2} R_{ll},\nonumber\\
& & R = -2R_{lklk}= 2R_{lk}.\label{e31}
\end{eqnarray}

Following Eq. (\ref{9}), the Ricci tensor can be written as,
\begin{eqnarray}
R_{\mu \nu} = \frac{1}{2 \alpha}[ \alpha R g_{\mu \nu} + g_{\mu \nu} f(T^\phi) +T^\phi_{\mu \nu} -2f_T\nabla_\mu \phi \nabla^\mu \phi]\label{51} 
\end{eqnarray} 

Using Eq. (\ref{e30}) and Eq. (\ref{e31}), one finds the following Ricci tensors:
\[ R_{lklk} \neq 0, R_{lml \bar{m}} \neq 0, R_{lklm} = R_{lkl \bar{m}} = 0. \]

From the above relation and Eq. (\ref{e30}), one finds the following NP quantities:
\[ \Psi_2 \neq 0 ; \Psi_3 = 0 ; \Psi_4 \neq 0,  \textrm{~~and~~}  \Phi_{22} \neq 0 \]

Thus we get four polarization modes for the GW: +,$ \times $ tensorial mode, breathing scalar mode and longitudinal scalar mode.

\section{Conclusion}\label{s5}

The theoretical foundations of modified gravity, a new approach intended to address and find solutions to the shortcomings and discrepancies of GR, are outlined in this report. These issues primarily manifest themselves at infrared and ultraviolet ranges, i.e., cosmological and astrophysical scales on the one hand and quantum scales on the other. 

The stability analysis of the scalar field varies depending on the circumstances of potential, and we have taken into account the spontaneous symmetry breaking analogous potential for our structure. The scalar field's behaviour varies identification and characterization of the critical parameter $( \mu^2 )$. The stable minimum value of the scalar field for $\mu^2 > 0$ is zero, resulting in a zero effective cosmological constant ($ \Lambda $). For $\mu^2 < 0$, the minimum scalar field would be non-zero, and the effective cosmological constant is non-zero as well. The variation of potential is shown in Fig. \ref{f2}. The $\mu^2 > 0$ variation is shown in red, whereas in blue coloured, the interpretation of $\mu^2 < 0$ is shown.

The scalar field Lagrangian is taken in conjunction may emerge a new set of Friedmann equations. Due to a mathematical constraint, the effective mass has a finite discontinuity. It is found for the range $ -0.50 \leq \beta \leq -0.25 $ effective mass is discontinuous. The variation is shown in Fig. \ref{f1}. 

The post-Minkowskian constraint of modified gravity, the problem of gravitational radiation, also deserves careful consideration.
When the gravitational action is just not Hilbert–Einstein, new polarizations emerge: in general, massive, massless, and ghost modes must be considered, whereas, in GR, only massless modes and two polarizations are present. This result necessitates a rethinking of GW physics. If GWs have nontensorial polarization modes, as mentioned, an analyzed signal, such as a stochastic cosmological background of GWs, would be an integration of each of these modes. 

In Einstein's General Relativity, the plus and cross modes of polarization are quite common. The plus mode is depicted by $ P_+=R_{txtx}+R_{tyty} $, the cross mode by $ P_{\times}=R_{txty} $, the vector-x mode by $ P_{xz}=R_{txtz} $, the vector-y mode by $ P_{yz}=R_{tytz} $, and the longitudinal mode by $ P_l=R_{tztz} $, and the transverse breathing mode by $ P_b=R_{txtx}+R_{tyty} $. 
For the form of potential $ V(\phi) = \frac{1}{2} \mu^2 \phi^2 + \frac{1}{4} \lambda \phi^4 $, in the frame of modified gravity $ f(R, T^\phi) = R + \beta T^\phi $, we obtain four polarization modes of GWs exists : +,$ \times $ tensorial mode, breathing scalar mode and longitudinal scalar mode, respectively.

\section*{Acknowledgements}
The southern federal university supported the work of SRC (SFedU) (grant no. P-VnGr/21-05-IF). SRC is also thankful to Ranjini Mondol of IISc, Bangalore, for the fruitful discussion to improve the manuscript. 

%% The bibliography section

\end{document}